\documentclass[aps,prx,twocolumn,floats,showpacs,superscriptaddress,nofootinbib]{revtex4-1}
\usepackage{graphicx,epsfig}% Include figure files
\usepackage[title,titletoc,toc]{appendix}
\usepackage{times,slashed}
\usepackage{dcolumn,bm,float}
\usepackage{graphics}
\usepackage{amssymb,amsmath,rotate,color}

\usepackage[colorlinks=true,citecolor=blue]{hyperref}

\newcommand{\be}{\begin{equation}}
\newcommand{\ee}{\end{equation}}
\newcommand{\bearr}{\begin{eqnarray}}
\newcommand{\eearr}{\end{eqnarray}}
\newcommand{\nn}{\nonumber}

\newcommand{\eps}{\varepsilon}
\newcommand{\vk}{{\mathbf k}}
\newcommand{\vp}{{\mathbf p}}
\newcommand{\vq}{{\mathbf q}}

\newcommand{\vQ}{{\mathbf Q}}

\newcommand{\bk}{{\mathbf k}}
\newcommand{\bp}{{\mathbf p}}
\newcommand{\bq}{{\mathbf q}}

\newcommand{\bs}{\boldsymbol}

\begin{document}

\title{Undamped inter-valley paramagnons in doped graphene}

\author{S. A. Jafari}
\affiliation{Department of Physics, Sharif University of Technology, Tehran 11155-9161, Iran}
\affiliation{Center of Excellence for Complex Systems and Condensed Matter (CSCM), Sharif University of Technology, Tehran 1458889694, Iran}
               
\author{J\"urgen K\"onig}
\affiliation{Theoretische Physik and CENIDE, Universit\"at Duisburg-Essen, 47048 Duisburg, Germany}

\date{\today}

\begin{abstract}
We predict the existence of an undamped collective spin excitation in doped graphene in the paramagnetic regime, referred to as \textit{paramagnons}. 
Since the electrons and the holes involved in this collective mode reside in different valleys of the band structure, the momentum of these inter-valley paramagnons is given by the separation of the valleys in momentum space.
The energy of the inter-valley paramagnons lies in the void region \textit{below} the continuum of inter-band single-particle electron-hole excitations that appears when graphene is doped.
The paramagnons are \textit{undamped} due to the lack of electron-hole excitations in this void region.
Their energy strongly depends on doping concentration, which can help to identify them in future experiments.
\end{abstract}

\maketitle
\section{Introduction}
Dirac semi-metals are conducting states of matter which are characterized by a linear band touching at isolated points in the Brillouin zone~\cite{Witten} with far reaching consequences~\cite{Neto,Wehling}.
Graphene is the most prominent example of such a Dirac material in two dimensions~\cite{Neto}.
States in Dirac materials are characterized by momentum, spin, and pseudospin.
The latter is an internal degree of freedom originating from the two orbital degrees of freedom that form the material
and is different from the physical spin.  
The effective theory of electronic states in graphene locks momentum and pseudospin to each other~\cite{GEIM2009},
while the physical spin simply comes in the free theory as a degeneracy factor. 
For positive-energy states, momentum and pseudospin are parallel to each other, giving rise to a (pseudospin) helicity (often referred to as chirality) of $\lambda=+1$.
Negative-energy states, on the other hand, have helicity $\lambda=-1$ (meaning that momentum and pseudospin are antiparallel to each other).
The chiral nature of Dirac states is the root of many fascinating properties of Dirac materials such as Klein tunneling~\cite{Katsnelson,Katsnelsonbook}.

Graphene hosts two Dirac cones per unit cell in momentum space, usually referred to as valleys~\cite{Yang2014,Katsnelsonbook}.  
Thus, in addition to the quantum numbers $k$ for momentum, $\sigma$ for spin, and $\lambda$ for helicity, there is a valley index $\tau=\pm 1$.
This leads to valley-polarized transport phenomena~\cite{Schaibley2016}, defining the field of valleytronics~\cite{Rycerz2007}.
For the issue of the present paper, namely the existence of undamped paramagnons in doped graphene, the valley degree of freedom is of crucial importance as it allows for collective excitations with electrons and holes residing in \textit{different} valleys, as indicated by the name \textit{inter-valley paramagnon}.

Interactions play an important role for the properties of electronic systems~\cite{vignale,senechal}.
Increasing Coulomb interaction can drive a phase transition from a paramagnet to a magnetically-ordered state.
This includes itinerant ferromagnetism~\cite{Lovesey} as described by the Stoner model but also antiferromagnetic 
N\'eel order in the Hubbard model for non-frustrated lattices at half filling~\cite{Fazekas,FradkinBook}.
In both cases, Goldstone's theorem ensures that the spontaneous breaking of spin symmetry yields a gapless branch of collective spin excitations, 
referred to as \textit{magnons}~\cite{Altland,Noorbakhsh}. Magnons are indicated by poles in the \textit{spin susceptibility}~\cite{Coleman}.
But also for magnetically disordered states, Coulomb interaction can give rise to collective excitations.
The most famous example is the \textit{plasmon}, a collective mode of charge-density excitations (carrying no spin), which is described by poles in the 
\textit{charge susceptibility}~\cite{vignale}.
Poles in the charge susceptibility may also appear for spin-polarized systems (such as a two-dimensional electron gas in a magnetic field).
The corresponding collective charge excitation has been dubbed \textit{spin plasmon}~\cite{PoliniSpinPlasmon}.

Table~\ref{collectives.tab} summarizes for the so-far mentioned collective excitations (magnon, plasmon, spin plasmon) the magnetic nature of the ground state to be excited (magnetically ordered/disordered) and how they are indicated (poles in the charge/spin susceptibility). 
Obviously, one combination is still missing to complete the scheme: a collective excitation indicated by poles in the spin susceptibility for a magnetically disordered state, referred to as \textit{paramagnons}.
Like magnons, they are spin excitations, indicated by poles in the spin susceptibility.
Unlike magnons they are not Goldstone modes since they appear in magnetically disordered (paramagnetic) states,  similar as plasmons do. 
Magnons and paramagnons admit a unified description in terms of a nonlinear sigma model~\cite{Villain1998}. 
The concept of paramagnons, i.e., collective spin excitations in magnetically disordered systems, has already been introduced more than 50 years ago in order to explain the reduction of superconducting pairing~\cite{Berk1966} and the increased effective electron mass~\cite{Doniach1966} in systems close to the Stoner instability. 
Paramagnons are somewhat related to collective excitations called \textit{triplons}.
The latter appear in strongly-correlated systems in which magnetic order is destroyed by \textit{quantum} fluctuations~\cite{Kohno2007}.
They have been interpreted as two-spinon bound states~\cite{Baskaran2003}, with spinons being complicated quasiparticles appearing in strongly-correlated systems.
Paramagnons, on the other hand, are collective excitations of systems in which magnetic order is destroyed by \textit{thermal} fluctuations, such that susceptibilities can be straightforwardly computed within an \textit{Random Phase Approximation} (RPA)-scheme directly in terms of \textit{electrons} and \textit{holes} rather than more complicated quasiparticles.

\begin{table}[t]
\begin{tabular}{|l|c|c|}
\hline
& disordered & ordered \\
& state & state \\
\hline
charge susceptibility& plasmon & spin plasmon\\ 
\hline
spin susceptibility & paramagnon & magnon\\
\hline
\end{tabular}
\caption{Collective excitations indicated by poles in the charge/spin susceptibility of a material in a magnetically ordered/disordered state.}
\label{collectives.tab}
\end{table}

The lifetime of collective excitations is limited by their decay into individual particle-hole (PH) excitations, referred to as \textit{damping}.
Outside the continuum of particle-hole excitation energies and momenta, however, there is no decay channel of the collective excitation into individual PH excitations.
As a consequence, the collective excitation remains \textit{undamped} which makes it experimentally accessible.
For many model systems, the paramagnons lie inside the PH continuum (PHC) and are, therefore, damped.
This is the reason why the term \textit{paramagnon} is sometimes used as a synonym for \textit{damped magnon}.
We, however, follow the convention that paramagnons denote collective spin excitations in a paramagnetic state, which can be either damped or undamped.
The main result of the present paper is to show that doped graphene does, due to its peculiar band structure, accommodate undamped paramagnons.

\begin{figure}[t]
\includegraphics[angle=0,width=.42\textwidth]{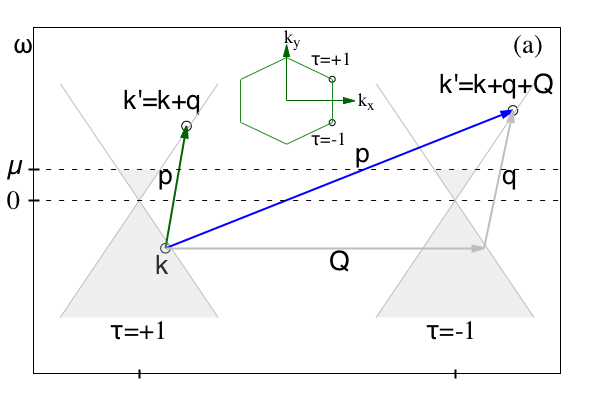}
\includegraphics[angle=0,width=.42\textwidth]{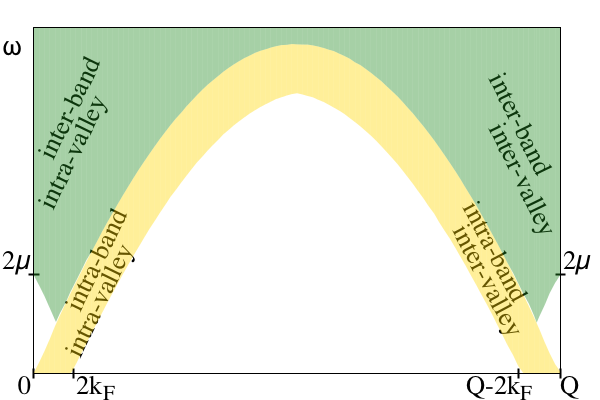}
\caption{
(a) Sketch of the band structure involving two independent Dirac cones, denoted by valley index $\tau=\pm1$.
The green (blue) arrow represents an intra- (inter-)valley excitation with momentum $\bp=\vq$ ($\bp=\vQ+\vq$), where $\vq$ denotes a small momentum.
(b) Schematic representation of the PHC. 
The green and gold shaded region correspond to inter-band ($\lambda\lambda'=-1$) and intra-band ($\lambda\lambda'=1$) PH processes, respectively.
The triangular void region for $\bp$ around $\vQ$ provides room for undamped inter-valley paramagnons.
}
\label{schematic.fig}
\end{figure}

\section{model and RPA resummation}

In order to simplify notation, we work in units in which $k_F=v_F=\hbar=1$.
This leads to natural units $L=k_F^{-1}$, $T=k_F^{-1}v_F^{-1}$, and $M=\hbar k_F v_F^{-1}$ for length, time and mass, respectively~\cite{Desloge1994}. 
We choose the coordinate system such that the graphene sheet lies in the $x-y$-plane and that the wavevector $\vQ$ connecting the two independent Dirac cones in the Brillouin zone is directed along the $y$-axis, see inset of Fig.~\ref{schematic.fig}~(a).
For this choice, the valley denoted by valley index $\tau=\pm 1$ is, for each physical spin $s=\uparrow, \downarrow$, described by the Hamiltonian
\be
   H_\tau=\sigma_x k_x+\tau\sigma_y k_y,
   \label{dirac.eqn}
\ee
where the $\sigma_{x,y}$ are Pauli matrices in pseudospin space, and $\vk = (k_x,k_y)$ is the wavevector relative to the node of the respective valley. 
The eigenenergies and corresponding eigenstates are given by $\eps_{\tau\lambda \bk}=\lambda k$ and
\be
   |\tau \lambda \vk\rangle = \frac{1}{\sqrt{2}} \begin{pmatrix}1 \\ \lambda \exp(i\tau\phi) \end{pmatrix},
   \label{eigen.eqn}
\ee
respectively.
Here, $\lambda=\pm 1$ is the helicity index, $k=|\bk|=\sqrt{k_x^2+k_y^2}$ the length of the wave vector $\vk$, and $\phi$ the angle between $\vk$ and the $x$-axis.
The pseudospin part $ f^{\lambda\lambda'}_{\tau\tau'}(\vk,\vk') =| \langle \tau \lambda \vk | \tau' \lambda' \vk' \rangle  |^2$ of the 
overlap between two wavefunctions is given by
\be
   f^{\lambda\lambda'}_{\tau\tau'}(\vk,\vk') = \frac{1+\lambda\lambda' \cos(\phi-\tau\tau'\phi')}{2}.
   \label{formfactor.eqn}
\ee
It is very important to take this overlap factor into account in the calculation of the collective excitations.
Approximating the overlap factor $f$ by $1$ within a simplified treatment would not only quantitatively affect the energy and the momentum of the paramagnon,
but more severely, it would completely ignore an important qualitative difference between intra- and inter-valley transitions.
The intra-valley ($\tau\tau'=1$) overlap factor depends on the difference $\phi-\phi'$, while for inter-valley ($\tau\tau'=-1$) case, the sum $\phi+\phi'$ enters instead.
This {\it phase flip}, which can be traced back to the fact that the two Dirac theories around the two valleys correspond to two different (but related) representations of the Clifford algebra (see the appendix), supports the formation of inter-valley paramagnons, whereas intra-valley paramagnons are strongly suppressed, as we will discuss below.

\begin{figure}[t]
\includegraphics[angle=0,width=.44\textwidth]{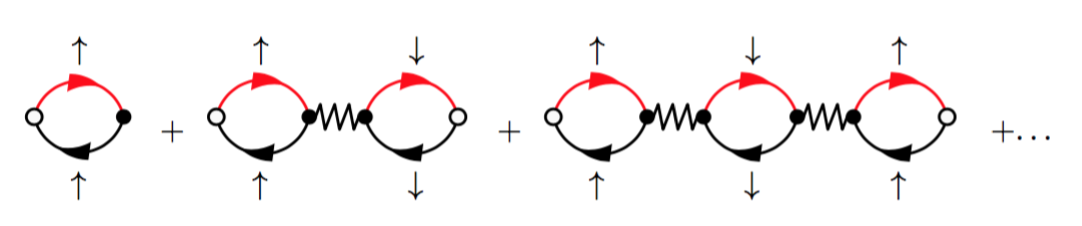}
\caption{ (Color online) RPA series for the susceptibilities.
Black and red lines represent fermion propagators in valley $\tau$ and $\tau'$, respectively.
The wiggly lines describe Hubbard-like Coulomb interaction, which is independent of momentum and only couples electrons of opposite spin.
The two vertices of each bubble combine to the overlap factor given in Eq.~\eqref{formfactor.eqn}.
For the charge susceptibility, the external vertices (open circles) are identical to the internal ones (filled circles).
For the longitudinal spin susceptibility, however, the external vertices introduce an extra factor $\sigma \sigma' (\hbar/2)^2$, where $\sigma$ and $\sigma'$ denote the spin of the first and the last bubble, respectively. 
}
\label{diagram.fig}
\end{figure}

In the absence of Coulomb interaction, both the charge and the spin susceptibility are governed by single-particle electron-hole excitations.
Diagrammatically, these processes can be represented by a single bubble diagram, as depicted by the first term of the diagrams shown in Fig.~\ref{diagram.fig}.
Attributing valley indices $\tau$ and $\tau'$ to the lower (black) and upper (red) lines for the particle and hole propagator, respectively, 
yields the general free susceptibility matrix
\be
   \chi^0_{\tau\tau'} (\bp, \omega) = \frac{1}{A}\sum_{\bk\lambda\lambda'} \frac{n_{\lambda \bk}-n_{\lambda' \bk'}}
   {\omega + \lambda k-\lambda' k' + i0^+} f^{\lambda\lambda'}_{\tau\tau'}(\vk,\vk')
   \label{chi0.eqn}
\ee
for energy $\omega$ and wavevector $\bp$. Therefore $\vp$ is the momentum of the particle-hole pair (bubble). 
The index $0$ indicates the absence of Coulomb interaction, and the sum runs over the band indices $\lambda$ and $\lambda'$ as well as over the 
initial wavevector $\vk$ of the hole. Furthermore, $A$ is the area of the graphene sample and $n_{\lambda k}$ is the occupation probability of 
state $|\tau \lambda \vk\rangle$, Eq.~\eqref{eigen.eqn} (which, at zero temperature, is either $0$ or $1$).
Both the initial and the final wavevector, $\vk$ and $\vk'$, describe deviations from the respective Dirac cone.
By introducing the wavevector $\vQ$ that connects the $\tau=+$ with the $\tau=-$ valley, we can write $\vk'=\vk+\vp$ where
the momentum of the particle-hole pair $\vp$ is parametrized in terms of an auxiliary momentum $\vq$ as 
\be
	\vp = \left\{ \begin{array}{ll} 
	\bq  & \quad \tau \tau' =1\\
	\bq + \tau \vQ & \quad \tau \tau' = -1
	\end{array} \right. 
	\label{kprime.eqn}
\ee
that guarantees that $\vq$ remains small for both intra- and inter-valley processes, see Fig.~\ref{schematic.fig}~(a).

In undoped graphene, the Fermi energy is at the nodes of the Dirac cones, $\mu=k_F=0$.
At zero temperature, all the states with $\lambda=-1$ are filled while the ones with $\lambda=1$ are empty.
Finite doping (either via chemical doping or via a gate voltage) introduces a finite Fermi energy $\mu$ (in this paper we choose $\mu>0$) and Fermi wavevector $k_F=\mu$.

The existence of an individual particle-hole excitation with wavevector $\bp$  and energy $\omega$ is indicated by a finite imaginary part of $\chi^0_{\tau\tau'} (\bp, \omega)$.
Intra- and inter-band contributions are marked as golden and green shaded areas in Fig.~\ref{schematic.fig}~(b), respectively.
Finite Coulomb interaction can support the formation of collective excitations, which remain undamped outside the particle-hole continuum.
In Refs.~\cite{Wunsch,Sarma2007}, it has been demonstrated that doped graphene can accommodate undamped collective charge excitations, i.e., plasmons.
Their dispersion has been derived from the charge susceptibility calculated within a random phase approximation (RPA) scheme.
Its diagrammatic representation consists of a series of bubbles, as shown in Fig.~\ref{diagram.fig}.
The black and red lines represent fermion propagators in valley $\tau=+$ and $\tau=-$, respectively.
The wiggly lines describe Coulomb interaction, which, in the following, we model by a Hubbard term.
Since the latter is local in space, it is independent of momentum transfer $\bp$ and, therefore, the same for intra-valley and inter-valley processes. 
Furthermore, the interaction line connects only electrons of opposite spin.
The RPA summation of the charge susceptibility gives~\cite{vignale,Coleman}
\be
   \chi_\text{charge}(\bp,\omega)= \frac{\chi^0 (\bp,\omega)}{1-U\chi^0 (\bp,\omega)}
   \label{chargeRPA.eqn}
\ee
for given valley indices $\tau$ and $\tau'$, which we dropped for keeping the equation transparent. 
The reason for the indices $\tau$ and $\tau'$ remaining the same in all bubbles of the RPA series is that valley-diagonal and valley-off-diagonal 
particle-hole processes are a vector $|\vQ|$ apart in the momentum space, see Eq.~\eqref{kprime.eqn}. 
The zeros of the denominator determine the energy-momentum relation of the plasmons.

The very same diagram series can be employed for the longitudinal spin susceptibility.
The only difference in the evaluation is the extra factor $\sigma \sigma' (\hbar/2)^2$, where $\sigma$ and $\sigma'$ denote the spin of the first and the last bubble, respectively. 
This yields 
\be
   \chi_\text{spin}(\bq,\omega)=\frac{\hbar^2}{4} \frac{\chi^0(\bq,\omega)}{1+U\chi^0(\bq,\omega)},
   \label{spinRPA.eqn}
\ee
which differs from the charge susceptibility in two respects.
First, there is an overall factor of $\hbar^2/4$.
Second and more important, all diagrams with an even number of bubbles acquire an extra minus sign~\cite{Scalapino}.
This alternating sign in the series is the reason for the different denominators in Eq.~(\ref{chargeRPA.eqn}) and (\ref{spinRPA.eqn}).
As a consequence of the different denominators for the charge and the spin susceptibility, undamped collective charge excitations reside \textit{above} the PHC while the undamped collective spin excitations lies \textit{below} it. 

The RPA series for the inter-valley and intra-valley processes are independent of each other.
The intra-valley processes involving a single Dirac cone has been studied by many authors~\cite{Sarma2007,Wunsch,Son2007}.
Therefore, we focus, in the rest of the paper on the inter-valley particle-hole propagator that dominates for $\bp$ around $\vQ$, such that $\chi^0(\bp,\omega) = \chi^0_{+-}(\bq+\vQ,\omega)$. 
Undamped paramagnons are indicated by poles of the spin susceptibility.
This leads to the condition
\be
	1+U \text{Re} \, \chi^0_{+-}(\bp,\omega) = 0 \, \, \, \, \text{and} \, \, \, \, \text{Im} \, \chi^0_{+-}(\bp,\omega) = 0 \, .
	\label{poleparamagnon.eqn}
\ee
The condition $\text{Im} \, \chi^0_{+-}(\bp,\omega) = 0$ is satisfied in the void region outside the PHC on the right hand side of Fig.~\ref{schematic.fig}~(b).
This corresponds to region B3 in Fig.~\ref{regions.fig}. 

\section{Undamped inter-valley paramagnons}

\begin{figure}[t]
\includegraphics[angle=0,width=.40\textwidth]{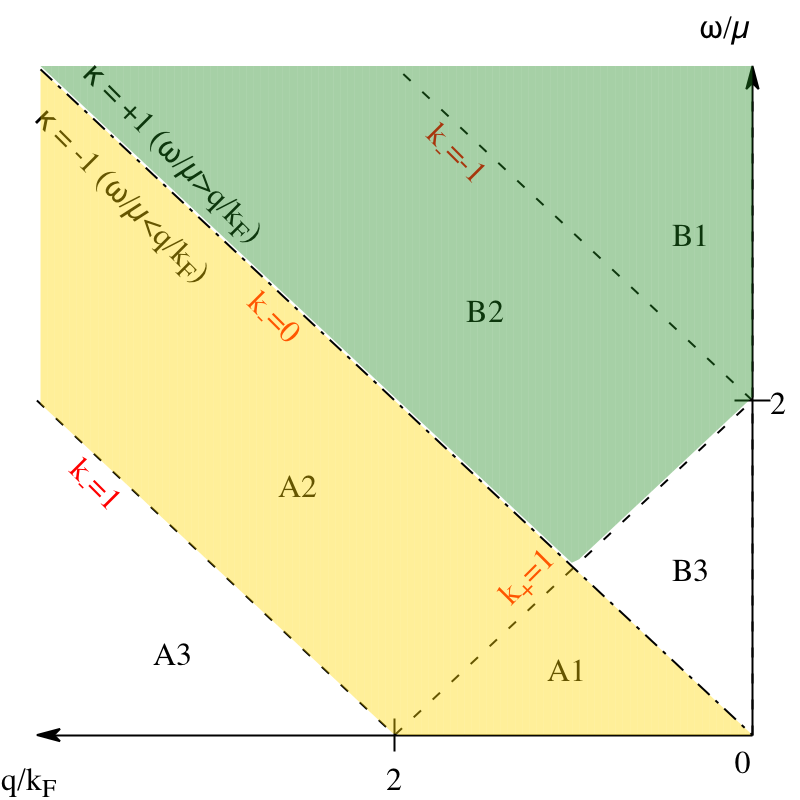}
\caption{ (Color online) The regions in the $(\omega,|\bq|)$ space relevant to the calculation of
the inter-valley polarization in doped graphene. The origin of $\vq$ corresponds to bubble momentum equal to $\vQ$. 
Dot-dashed line divides between $\omega<|\bq|$ (A-family) and
$\omega>|\bq|$ (B-family) corresponding to sign factor $\kappa=\mp 1$ in Eq.~\eqref{contourint.eqn}, respectively. 
Parameters $k_\pm$ of Eq.~\eqref{kpm.eqn} appearing in Eqs.~\eqref{gA.eqn} and~\eqref{gB.eqn} label the boundaries as indicated by red labels. 
Regions A1, A2 (gold filled) correspond to intra-band PHC, while B1, B2 (green filled) correspond to inter-band PHC. 
Regions A3, B3 are void of free PH excitations. 
The width of gold stripe and height of B3 are controlled by $k_F$ and $\mu$ and shrink to zero in the absence of doping. 
}
\label{regions.fig}
\end{figure}

We need to calculate the inter-valley susceptibility $\chi^0_{+-}$ as a function of $\omega$ and $\vq = \bp -\vQ$. 
Our calculation follows along the lines of Ref.~\cite{Wunsch}, in which the intra-valley susceptibility $\chi^0_{++}$ has been meticulously calculated.
First, we define 
\be
   \chi^0_{+-}=\Delta\chi^0_{+-}+\chi^0_{u,+-},
   \label{twoterms.eqn}
\ee
where $\chi^0_{u,+-}$ is the contribution to the inter-band, non-interacting spin susceptibility of undoped graphene. At zero temperature, the occupation numbers are 
either $0$ or $1$. 
If we subtract the undoped part $\chi^0_{u,+-}$ and perform a change of variables $\vk+\vq\to\vk$, then the occupancy of $1$ contributes only for $k<1\equiv k_F$. 
This yields
\be
  \Delta\chi^0_{+-}=\frac{1}{4\pi^2}\sum_{\alpha,\lambda,\lambda' }\int'd^2\vk
  \frac{f^{\lambda \lambda'}_{+-}(\vk,\vk') }{\alpha \omega_++(k-\lambda \lambda' k')},
  \label{deltachi01.eqn}
\ee
where the prime on the integral indicates the condition $k < 1$.
The extra summation over $\alpha=\pm 1$ emerges from a change of variables that 
converts the restriction on $k'$ into a restriction on $k$. The above integral can be
analytically calculated. The details of calculations are given in the appendix~\ref{calc.sec}. 
The end result is
\be
   \Delta\chi^0_{+-}=-\frac{1}{2\pi}\frac{\omega^2}{q^2}-\frac{1}{8\pi}\frac{\omega^2}{|\omega^2-q^2|}
   \left\{\begin{array}{lr}
      F_A	&\mbox{A regions}\\
      F_B	&\mbox{B regions}
   \end{array}\right.
   \label{delchiinter.eqn}
\ee
where the small momentum $\vq$ is measured from the other valley and 
the functions $F_A(\omega,\bq)$ and $F_B(\omega,\bq)$ are given in Eqs.~\eqref{gA.eqn} and~\eqref{gB.eqn}, respectively.
The A, B regions are depicted in Fig.~\ref{regions.fig}. 
The corresponding integration for the undoped term, $\chi^0_{u,+-}$ can be obtained by dimensional regularization
involving two different representations of Dirac matrices (see appendix~\ref{interundoped.sec}) and becomes
\be
   \chi^0_{u,+-}=\frac{1}{16}\frac{\omega^2-3\bq^2/2}{\sqrt{v_F^2\bq^2-\omega^2}}.
   \label{chi0JafariKoenig.eqn}
\ee
Adding Eq.~\eqref{chi0JafariKoenig.eqn} and~\eqref{delchiinter.eqn} according to~\eqref{twoterms.eqn}
will give the total inter-valley PH propagator in doped graphene. 

Eqs.~\eqref{delchiinter.eqn} and~\eqref{chi0JafariKoenig.eqn} are main technical results of this paper.
The details of derivation are given in the appendix. 
In order to use it properly, we consider the $(\omega,\bq)$ space shown in Fig.~\ref{regions.fig}.
The origin of $\bq$ is fixed at the $\vQ$ that corresponds to the wavevector connecting the two valleys $\tau=+$ and $\tau=-$.
The momentum of the particle-hole pair (bubble) is $\vq+\vQ$ and hence $\bq=0$ excitations actually have
a very large center of mass momentum equal to the wave vector $\vQ$.
The inverted direction of $\bq$ in Fig.~\ref{regions.fig} is meant to emphasize that it corresponds to the corresponding 
triangular region at the bottom right of Fig.~\ref{schematic.fig}. 
The dot-dashed line bisecting the $(\omega,q)$ plane separates the A regions ($\omega<q$) from B regions ($\omega>q$).
The location of the step in $\Theta$ functions of~\eqref{gA.eqn} and~\eqref{gB.eqn}
are given by 
\be
   k_\alpha=\frac{q+\alpha\omega}{2},~~~~~~\alpha=\pm
   \label{kpm.eqn}
\ee
which define the dashed curves in Fig.~\ref{regions.fig}. 
The gradient of $k_-$ is perpendicular to the bisector of $(\omega,q)$, heading towards $q$ axis. 
The gradient of $k_+$ is along the bisector and points away from the origin of $(\omega,q)$ coordinates. 
For example, the B3 region of interest for our paramagnon mode, is characterized by $k_+<1$ {\em and} $-1<k_-<0$. 
Putting the above conditions together gives, 
\be
   B3:~\Theta(1-k_+)\Theta(1+k_-)\Theta(-k_-)
   \label{B3region.eqn}
\ee
which has intersections with the first three terms of Eq.~\eqref{gB.eqn}.  

Regions A1, A2 in Fig.~\ref{regions.fig} correspond to inter-valley PH excitations 
that take place across the Fermi surface, i.e. from the interior of the Fermi surface in one valley to the
exterior of the Fermi surface in the other valley which is depicted in panel (a) of Fig.~\ref{schematic.fig}. That is why
the width of the gold strip in Fig.~\ref{regions.fig} is controlled by $2k_F$ (which in natural units is $2$). 
The spectral density of the PH propagator coming from the golden region is responsible
for the inter-valley plasmons~\cite{MikhailovIntervalleyPlasmon}.
The regions B1 and B2 (green shaded) correspond to inter-band PH excitations. These are PH excitations that 
leave the hole in the negative energy branch of Dirac cone ($\lambda=-1$) and the electron in the positive energy branch ($\lambda=+1$)
of the {\it other} Dirac cone. The spectral density of the PH propagator arising from such inter-band excitations
corresponding to regions B1 and B2, plays an essential role in formation of the paramagnon pole. 

\begin{figure}[t]
\includegraphics[angle=0,width=.49\textwidth]{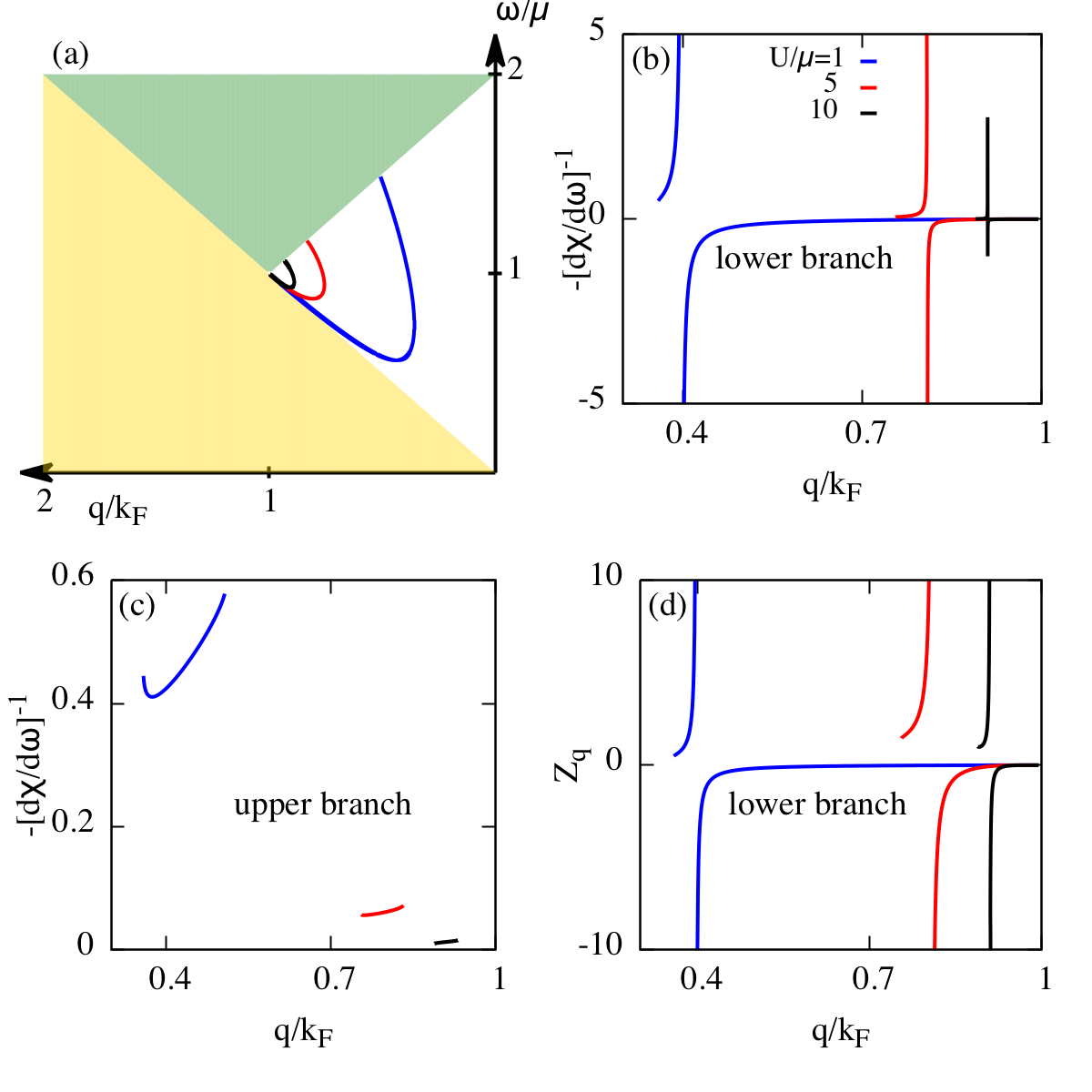}
\caption{(Color online) Position of the paramagnon poles of the inter-valley RPA susceptibility in the B3 region. 
The blue, red, and black curves corresponds $U/\mu=1,5,10$, respectively. (a) denotes the dispersion of
paramagnon poles. The inverse slope $\partial \chi^0_{+-}/\partial \omega$ at the lower and upper poles are plotted in panels (b) and (c).
Panel (d) shows the same set of data as in (b) but multiplied with the factor $U^2$ to yield the pole strength, see Eq.~\eqref{residue.eqn}. 
This shows that the coupling of the paramagnon pole to a neutron probe is jointly determined by the above derivative (coming from the
phase space of PH states) and the interaction parameter $U$. 
For the above values of $U$, the typical strength of the pole is $\sim 10^2$ which is a very strong pole as compared to Fermionic residues which are always less than one. 
}
\label{B3solution.fig}
\end{figure}

We are now ready to decipher Eq.~\eqref{poleparamagnon.eqn} in region B3. 
In this region, the ${\rm Im}\chi^0_{+-}$ is zero. So one only needs to solve the real part in Eq.~\eqref{poleparamagnon.eqn}.
Since the $\chi^0_{u,+-}$ portion in Eq.~\eqref{chi0JafariKoenig.eqn} will not contribute to real part, it plays no role in determination 
of the location of the poles. 
The triangular region B3 in Fig.~\ref{regions.fig} as pointed out in Eq.~\eqref{B3region.eqn} has
intersections with the first three terms of Eq.~\eqref{gB.eqn}. 
But from Eq.~\eqref{evalg.eqn}, the value of function $g_B$ (i.e. $g_{\kappa=+1}$) 
%--- $\kappa=+1(-1)$ essentially means B (A) regions. See below Eq.~\eqref{contourint.eqn} ---
at $x=-1$ gives $\ln (-1)$.
Furthermore in region B3, the argument $-\omega/q$ of the second $g_B$ function is always less than $-1$, and hence the logarithm
in the definition of $g_B$ contributes another minus sign which cancels the other $\ln(-1)$, and therefore only the $x\sqrt{x^2-1}$ part of it contributes.
Therefore, Eq.~\eqref{poleparamagnon.eqn} reduces to,
\bearr
   &&1=\\
   &&\frac{\omega^2}{8\pi}U\left\{\frac{4}{q^2}+\frac{1}{|\omega^2-q^2|}\left[g_B\left(\frac{2-\omega}{q}\right)
   -\frac{\omega}{q}\frac{\sqrt{\omega^2-q^2}}{q}\right]\right\}. \nn
\eearr
Numerical solutions of this equation for three values of $U/\mu=1,5,10$ are plotted in panel (a) of Fig.~\ref{B3solution.fig}
and correspond to blue, red and black dispersions, respectively. As can be seen, there are lower and upper branches. Near the
pole, the RPA susceptibility behaves like $Z_{\bq}/(\omega-\omega_{\rm pm}(\bq))$, where $\omega_{\rm pm}(\bq)$ denotes either of the
lower or upper paramagnon branches. The pole strength 
\be
   Z_{\bq}=-\left[\left.U^2\partial \chi^0_{+-}(\omega,\bq)/\partial \omega\right|_{\omega_{\rm pm}(\bq)}\right]^{-1},
   \label{residue.eqn}
\ee
governs the neutron scattering signal~\cite{Jafari2005}.
In panels (b) and (c) of Fig.~\ref{B3solution.fig}, we have plotted this quantity without inclusion of $U$
to emphasize the phase space related aspect of the pole strength. In panel (d), we have included $U$
to plot the above $Z_q$ for the lower branch to emphasize the role of $U$ in determining the strength of the pole. 
As can be seen in panel (c), the strength of the pole at the
upper paramagnon branch is $10^{-1}$-$1$, while on the lower branch in panel (b) it varies from $\sim 3$ (black curve, $U=10\mu$)
to few tens (red curve, $U=5\mu$) and even few hundreds (blue curve, $U=\mu$).
Therefore for smaller $U/\mu$, the dispersion of the paramagnon occupies larger momentum region, and
the strength of the pole also increase. 
The comparison of the pole strengths for the lower and upper branches shows that for practical purposes,
the dominant coupling to a neutron probe comes from the lower branch and the neutron signal will get weaker upon 
passing through the turning point. 

It should be emphasized that the wave vector $\vq$ in this figure is pointed towards the other valley.
In actual graphene, there are three such directions connecting a valley with other valleys (due to $C_3$ symmetry). 
Away from these three directions, the overlap factors $f$ of Eq.~\eqref{formfactor.eqn} reduce the density of
PH pairs contributing to $\chi^0_{+-}$, and therefore the solutions gets faint upon deviations from the above
three directions.

\section{Summary and Discussions}
We have established the existence of an undamped paramagnon pole in doped graphene which arises from inter-valley 
processes. The corresponding bubbles are denoted with two colors in Fig.~\ref{diagram.fig} to emphasize that 
the electron and hole in the electron-hole propagator belong to two different valleys. The spinor phase flip
arising from a change in the color at the vertex, given by Eq.~\eqref{formfactor.eqn}, is responsible for the formation of the 
inter-valley paramagnon pole. The same phase flip is associated with the fact that the representation of the Dirac matrices
for the two valleys are related by complex conjugation (time-reversal) operation (see appendix). 
The decisive factor controlling the dispersion of the inter-valley paramagnon
is the ratio of the Hubbard $U$ to doping level $\mu$. The latter can be conveniently tuned in graphene across
orders of magnitude. Therefore, the energy scale $\sim\mu$ of the paramagnon branch can be controlled by the gate voltage. 

The RPA analysis of the longitudinal spin susceptibility revealed undamped paramagnons at large momentum transfer $\vp \approx \vQ$.
This triggers the question whether paramagnons do also exist at small momenta, $\vp\approx \mathbf{0}$?
In the case of momentum transfer $\vp \approx \vQ$, electron-hole fluctuations are dominated by the inter-valley contribution $\chi^0(\bp,\omega) = \chi^0_{+-}(\bp,\omega)$.
For small momenta, however, only intra-valley contributions survive, $\chi^0(\bp,\omega) = \sum_\tau\chi^0_{\tau\tau}(\bp,\omega)$.
It turns out that the overlap factors are very different for the two cases.
To see this, 
let us look at inter-band transitions, i.e., when the electrons from negative-energy states are excited to positive-energy states.
This implies $\lambda \lambda'=-1$.
Inter-valley transitions, which are relevant at large momentum transfer correspond to $\tau \tau'=-1$.
For momentum transfer close to $\vQ$, we get $\phi \approx \phi'$, which leads to the overlap factor $[1-\cos (2\phi)]/2$.
It vanishes for $\vk \perp \vQ$ but is unity for $\vk \parallel \vQ$.
As a consequence, some of the fluctuations involving inter-valley (inter-band) transitions are suppressed but some are not.
The latter are responsible for the formation of paramagnons.
The situation is distinctively different for intra-valley transitions, $\tau \tau'=1$.
At small momentum transfer, again we get $\phi \approx \phi'$, and the overlap factor $[1-\cos (\phi-\phi')]/2$ vanishes.
This leads to smaller phase space for the processes involving intra-valley (inter-band) transitions.
As a result, the RPA analysis of the spin susceptibility delivers an undamped intra-valley collective mode only at very high values of the 
Coulomb interaction $U$ that are unrealistic for graphene.
This reasoning does not strictly prove the non-existence of undamped intra-valley paramagnons at small momentum transfer. 
Whether a resummation of another class of diagrams than those included in RPA may give yield paramagnons near the $\Gamma$ point or not remains an open question. 

We remark that for paramagnetic (i.e. magnetically disordered) states, transverse and longitudinal spin susceptibilities should contain identical information.
According to this, the inter-valley paramagnon should also be indicated by the transverse spin susceptibility.
The latter, however, cannot be described by the series of bubble diagrams discussed above. 

In the region A3, which exists in both doped and undoped graphene, there is no distinction between inter-valley 
and intra-valley processes, and the entire band structure contributes to the value of the bubble diagram. 
Earlier numerical works indicate a paramagnon pole~\cite{EbrahimkhasIJMPB,Jafari2005,Jafari2002,Jafari2004,Posvyanskiy} in
region A3 as well. 
An inter-valley {\em plasmon}, lending on the contributions from A1 and A2 regions was discussed earlier~\cite{MikhailovIntervalleyPlasmon}.
However, the meticulous calculation in the present work, additionally includes the contribution of regions B1 and B2 as well,
which eventually generates the paramagnon poles. A quite analogous mode has been found in the doped Dirac cone at the surface of
topological insulators~\cite{Maslov2017} where the appearance of {\em physical spin} (rather than the sublattice pseudo-spin)
has a similar spinor phase flip effect appearing in the inter-valley PH excitations. 

What are possible contributions of such a paramagnon mode to anomalies in graphene? 
The existence of a paramagnon mode in a substantial portion of the Brillouin zone clearly 
indicates that a spin-1 boson exhausts energy scales from $\sim\mu$ all the way up to the hopping energy 
scale $\sim 2$eV. The doping levels routinely available in graphene can be as large as $\mu\sim 0.5$ eV. 
For unscreened Hubbard $U$ values of few eV, the ratio $U/\mu$ can easily exceed $10$ (black curve in Fig.~\ref{B3solution.fig}).
For smaller values of $\mu$, the ratio $U/\mu$ will be further enhanced,
thereby shrinking the undamped mode in B3 region into a point, reminiscent of the $41$ meV neutron scattering peak in YBCO
superconductor~\cite{Aeppli1993,Keimer1997}. Note that in our model with two cones, there is only one such spot.
However, {\em in realistic graphene, there will be three of them related by $C_3$ rotations.} 
In the following, we list some possible effects arising from a paramagnon excitation: 
(i) The fact that paramagnons do not carry electric charge (as it is formed by electron and hole pairs)
will contribute to the violation of Wiedemann-Franz law~\cite{Crossno2016,Seo2017}. 
(ii) The very basic vertex describing the coupling of paramagnons to fermions may act as a separate
source of spin current noise~\cite{Matsuo2018}. It further suggests the extension of the
standard hydrodynamic description of the Dirac electron fluid~\cite{Lucas2018} by inclusion of 
a novel {\em spin viscosity}.
(iii) The fact that the energy of this mode is $\sim \mu$ is expected to play a role in the
anomalous optical absorption of graphene~\cite{Basov2008} at energies $\sim 2\mu$ for which a free
Dirac theory can not account. (iv) Last, but not least, in recent experiments on Li-doped graphene, the 
inter-valley scattering rates have been extracted from the total scattering rates~\cite{Khademi2019}. The 
theoretical prediction based on the scattering of {\em free} Dirac electrons from impurities fails to reproduce
the entire experimentally observed scattering rates. Our paramagnons whose energy is set by the
chemical potential can serve as a possible decay channel. Earlier scattering rates extracted from
ARPES data on hydrogenated graphene also indicate that the scattering of free electrons from 
impurities is not able to account for the observed scattering rate~\cite{haberer2011electronic}. Again our paramagnon branch 
can serve as an additional scattering channel. 

\section{acknowledgments}
S. A. J. was supported by Alexander von Humboldt fellowship for experienced researchers and
grant No. G960214 from the research deputy of Sharif University of Technology and also 
Iran Science Elites Federation (ISEF). We acknowledge insightful discussions with G. Baskaran.

% - - - - - - - - - - - - - - - - - - - - - - - - - - - - - - - - - - - - - -
\appendix
\section{Calculation of inter-valley polarizability}
\label{calc.sec}
Since in the literature there is no calculation related to the valley-off-diagonal 
component of the polarization function in graphene, we give the full details of the 
calculation of $\chi^0_{+-}$ to enable the reader to verify all the steps. 
As pointed out in Eq.~\eqref{twoterms.eqn}, the inter-valley susceptibility breaks into two contributions.
In this appendix, we calculate the first term of~\eqref{twoterms.eqn}, namely $\Delta\chi^0_{+-}$, 
and in appendix~\ref{interundoped.sec} we calculate the second term, namely $\chi^0_{u,+-}$ for undoped graphene, 
using dimensional regularization and field theory methods. 

Starting from Eq.~\eqref{deltachi01.eqn} and performing the summation over $\zeta$ gives,
\bearr
&&\Delta\chi^0_{+-}=\int'\frac{d^2\vk}{4\pi^2}\sum_{\alpha=\pm1}
\frac{(k+\alpha\omega_+)+k' C_{\vk,\vk'}}{(k+\alpha\omega_+)^2-k'^2}
\eearr
where $C_{\vk,\vk'}=\cos(\phi_{\vk}+\phi_{\vk'})$. Taking $\vq=\vk'-\vk$ along
the $x$-axis, and assuming that the angle between $\vk$ and $x$-axis is $\varphi$ one can nicely 
obtain $k'C_{\vk,\vk'}=\left(q\cos\varphi+k\cos2\varphi\right)$. If the angle $\theta$ subtended by $\vq$ and the $x$-axis 
was nonzero, the $q\cos\varphi$ in this expression would be replaced by $q\cos(\varphi+\theta)$. 
Therefore,
\bearr
&&\Delta\chi^0_{+-}=\int'\frac{kdkd\varphi}{4\pi^2}\sum_{\alpha=\pm1}
\frac{(k+\alpha\omega_+)+q\cos\varphi+k\cos2\varphi}{\omega_+^2+2\alpha k\omega_+-q^2-2kq\cos\varphi}\nn.
\eearr
Since $\alpha\omega_++q\cos\varphi+2k\cos^2\varphi=(t+\cos\varphi)[2k\cos\varphi+s]+r$ where,
\bearr
   t&=&\frac{\omega_+^2+2\alpha k\omega_+-q^2}{-2kq}\nn\\
   s&=&q-2kt=\frac{\alpha\omega_+}{q}(2k+\alpha\omega_+)\nn\\
   r&=&\alpha\omega_+ -ts\nn\\
   &=&\frac{\omega_+^2}{2kq^2}\left[(2k+\alpha\omega_+)^2-q^2\right]\nn
\eearr
we can immediately evaluate the $\varphi$-integral.  
Note that in the case of intra-valley processes, instead of $k\cos2\varphi$ we would have $k$
and hence the $s$ and $r$ terms for intra-valley processes would be quite different from the above values~\cite{Wunsch}. 
The $\cos\varphi$ integrates to zero, while the $\varphi$ integral multiplying the $s$ term simply 
gives a factor of $2\pi$. To proceed with the $r$ term, let us set $\omega_+\to\omega$ and
focus on the real part of $\Delta\chi^0_{+-}$. In this case $t$ will be real and we use
the standard contour-integration formula,
\be
   \int_0^{2\pi} \frac{d\varphi}{t+\cos\varphi}=\frac{2\pi~{\rm sgn}(t)\Theta(|t|-1)}{\sqrt{t^2-1}},
   \label{int.eqn}
\ee
which requires the evaluation of
\be
   \frac{1}{\sqrt{t^2-1}}=\frac{2kq}{\sqrt{(\omega^2-q^2)\kappa}}\frac{1}{\sqrt{[(2k+\alpha\omega)^2-q^2]\kappa}},
   \label{contourint.eqn}
\ee
where we have introduced a sign variable $\kappa=+1$ ($-1$) when $\omega>q$ ($\omega<q$) corresponding to 
regions B (A) in Fig.~\ref{regions.fig}. So $\kappa$ is essentially a label for regions A, B in this figure. 
Let us define $x=(2k+\alpha\omega)/q$ as new integration variable which will temporarily replace $k$ variable. 
The sign factor replaces $\omega^2-q^2$ under the square root by $|\omega^2-q^2|$ and places the
extra minus sign of the $\omega<q$ region (denoted by A1, A2, A3 in Fig.~\ref{regions.fig}) case 
in the second square root of Eq.~\eqref{contourint.eqn}. 
In terms of the auxiliary variable $x$, it turns out that $\omega^2\lessgtr q^2$ and $|x|\lessgtr 1$ have same meanings.

A very significant difference of the inter-valley susceptibility with respect to intra-valley and 
2DEG cases is the form of the $s$ term. In this term the summation over $\alpha=\pm 1$ eliminates the terms 
that are odd with respect to $\alpha$ and leaves
\be
    -\frac{1}{2\pi}\frac{\omega^2}{q^2}.
\ee
Finally the $r$ term gives,
\be
   \frac{-1}{8\pi}\sum_{\kappa,\alpha}\frac{\omega^2}{\sqrt{(\omega^2-q^2)\kappa}}\int'\left[
   {\rm sgn}(t)\Theta(|t|-1)dg_\kappa\right],\label{rterm.eqn}
\ee
where $\kappa$ labels regions A and B in Fig.~\ref{regions.fig} and 
$dg_\kappa=2\sqrt{(x^2-1)\kappa}~dx$, or equivalently
\be
   g_\kappa(x)=x\sqrt{(x^2-1)\kappa}-\left\{
   \begin{array}{lr}
   \pi/2-\sin^{-1}(x)			& \kappa=-1\\
   {\rm sgn}(x)\cosh^{-1}(|x|)		& \kappa=+1
   \end{array}\right. ,
   \label{evalg.eqn}
\ee
where $\cosh^{-1}(x)=\ln(x+\sqrt{x^2-1})$. The validity of the above integral can be verified by direct differentiation.
Note that since the derivative of $g$ is an even function of $x$, the $g$ itself must be odd up to
an additive integration constant. The $|x|>1$ ($\kappa=+1$) piece is manifestly odd. 
We choose the integration constant $\pi/2$ in the first piece ($\kappa=-1$ part)
to ensure that the function $g$ is continuous at $x=+1$. For later reference, in $\kappa=\pm$ regions
the function $g_\kappa$ at $x=1$ gives $g_\kappa(1)=0$. This will simplify the definite integrations.

\begin{figure}[t]
\includegraphics[angle=0,width=.48\textwidth]{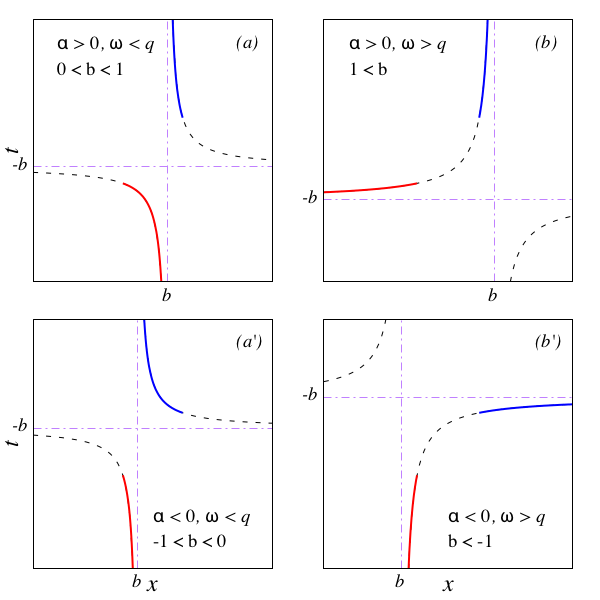}
\caption{
Plot of the function $t(x)=-b+\frac{1-b^2}{x-b}$ with $b=\alpha\omega/q$ for four different situations: 
(a) $\alpha>0$, $\omega>q$, (b) $\alpha>0$, $\omega<q$, (c) $\alpha<0$, $\omega>q$, (d) $\alpha<0$, $\omega<q$. 
The dashed parts of the curve correspond to $|t(x)|<1$ that do not contribute to the integral. 
The solid part that contribute with positive (negative) sign is denoted by blue (red) color
and ends at $x^+=+1$ ($x^-=-1$). The horizontal and vertical dot-dashed lines are asymptotes $x=+b$ and $t=-b$, respectively
}
\label{signs.fig}
\end{figure}

The remaining task is to carefully determine the conditional expressions arising
from the step function and the sign function in Eq.~\eqref{rterm.eqn} which are either zero or $\pm 1$
and multiply the $g_\kappa$ function of Eq.~\eqref{evalg.eqn}. In terms of the new variable $x$ we 
can rewrite variable $t$ as,
\bearr
   t=\frac{\alpha\omega x-q}{\alpha\omega-qx}=-b+\frac{1-b^2}{x-b},~~~b=\frac{\alpha\omega}{q}. 
\eearr
The critical values $t^\pm=\pm 1$ precisely correspond to $x^\pm=\pm 1$.
In Fig.~\ref{signs.fig} we have plotted $t$ as a function of $x$ for four possible situations $\alpha\lessgtr 0,~\omega\lessgtr q$.
The dashed curves correspond to $|t|<1$ which by step function in Eq.~\eqref{rterm.eqn} do not contribute. 
Only the solid portions of the curves correspond to non-zero contribution in Eq.~\eqref{rterm.eqn}. 
The sign is encoded into the color of solid line. 
The blue (red) corresponds to contribution with positive (negative) sign. The blue (red) lines
end at $x^+=+1$ ($x^-=-1$).

Let us examine the $\omega<q$ region in Fig.~\ref{signs.fig} that corresponds to panels (a) and (a').
The condition $\omega<q$ marks regions A1, A2, A3 in Fig.~\ref{regions.fig} and hence amounts to setting $\kappa=-1$ in Eq.~\eqref{contourint.eqn}.
In this case, irrespective of the sign of $\alpha$ in panels (a) and (a'), we find that the the sign of $t$ is positive (blue color in Fig.~\ref{signs.fig})
when $b<x<1$ and it is negative (red color) when $-1<x<b$.
Translating from variable $x$ to variable $k$, the region $t>1$ ($t<-1$) corresponds to $0<2k<q-\alpha\omega$ ($-q-\alpha\omega<2k<0$).
This gives the term $\Theta(-\kappa)\left[\Theta(k)\Theta(q-\alpha\omega-2k)-\Theta(-k)\Theta(2k+q+\alpha\omega)\right]$
where $\Theta(-\kappa)$ denotes the region A in Fig.~\ref{regions.fig}. But since $\Theta(-k)$ term
has no intersection with the Fermi sphere defined by $0<k<1$, the second term will not contribute to the
$k$-integral. So in region A, we are only left with,
\be
   \sum_{\alpha}\Theta(k)\Theta(q-\alpha\omega-2k),~~~\mbox{A regions}. 
   \label{thetaA.eqn}
\ee

Similarly setting $\kappa=+1$ in Eq.~\eqref{contourint.eqn} corresponds to regions B1, B2, B3 in Fig.~\ref{regions.fig} 
and hence the condition $q<\omega$ that labels panels (b) and (b') in Fig.~\ref{signs.fig}. 
In this case the result depends on $\alpha$: In panel (b), where $\alpha=+1$, 
the positive sign of $t$ (blue color in Fig.~\ref{signs.fig}) corresponds to $1<x<b$ and the negative sign (red curve) corresponds to
$x<-1$. These two conditions when translated from $x$ to $k$ will correspond to two conditions $q-\omega<2k<0$, and $2k<-q-\omega$, respectively. 
But these conditions have no intersection with the Fermi sphere $0<k<1$ and hence no
$\Theta(\alpha)$ contribution arises. But in panel (b') where $\alpha=-1$ the positive (negative)
sign for $t$ is obtained for $x>1$ ($b<x<-1$) which in terms of $k$ becomes $2k>q+\omega$ ($0<2k<-q+\omega$). 
In this case, both regions may have intersection with the Fermi sphere, and therefore 
we collect the following result:
\be
   \left[\Theta(2k-q-\omega)-\Theta(\omega-q-2k)\Theta(k)\right],~~\mbox{B regions}.
   \label{thetaB.eqn}
\ee

Now the $\mbox{sgn}(t)$ values originated in Eq.~\eqref{int.eqn} and needed in Eq.~\eqref{rterm.eqn} are encoded as
positive or negative signs of appropriate $\Theta$ functions in Eqs.~\eqref{thetaA.eqn} and~\eqref{thetaB.eqn} and the remaining integral
over $k$ (or equivalently $x$) is expressed in terms of a complete differential $dg_\kappa$. So all we need to look into the 
intersection of the Fermi sphere $\Theta(k)\Theta(1-k)$ with the restrictions~\eqref{thetaA.eqn} or~\eqref{thetaB.eqn}
to figure out the limits of integration. The end result for the $r$ term will be the difference
$g_{\kappa}(x_{\rm max})-g_\kappa(x_{\rm min})$. 

In A region two values of $\alpha=\pm 1$ contribute. According to Eq.~\eqref{thetaA.eqn},
the upper limit of $k$ integration is given by $k_{-\alpha}$, where $k_\pm$ is defined in Eq.~\eqref{kpm.eqn}.
The lower limit of $k$ integral is always $0$, while the upper limit is ${\rm min}(1,k_{-\alpha})$. 
Noting that at $k_{\rm max}=k_{-\alpha}$ one has $x_{\rm max}=1$ and that $g_\kappa(1)=0$, 
after a $\alpha\to -\alpha$ under the $\sum_\alpha$ this leads to 
\be
   F_A=\sum_\alpha g_A\left(\frac{2-\alpha\omega}{q}\right)\Theta\left(1-k_{\alpha}\right)
   -g\left(\frac{-\alpha\omega}{q}\right),
   \label{gA.eqn}
\ee
where $g_A$ is obtained from Eq.~\eqref{evalg.eqn} by setting $\kappa=-1$. 

Similarly for the B regions we need to define a lower limit for $k$ in the first
term of Eq.~\eqref{thetaB.eqn} which turns out to be the same as $k_+$.
Since $q$ and $\omega$ are both positive, $k_+$ is always positive. So we only need to determine its location 
with respect to the radius of Fermi sphere, $1$. The only possible way for the first term
to be nonzero is $k_+<1$ such that the integrals over $k$ extends from $k_+$ to $k=1$
that correspond to the range from $x_{\rm min}=1$ to $x_{\rm max}=(2-\omega)/q$. So the first
term gives $\Theta(1-k_+) g_{B}\left((2-\omega)/{q}\right)$. 
From the second term of Eq.~\eqref{thetaB.eqn} which carries an overall minus sign from the $\rm{sgn}(t)$, 
the $k_{\rm max}$ turns out to be $-k_-$. 
In this case, the lower limit of $k$ integration is always zero, giving $x_{\rm min}=-\omega/q$. 
But the upper limit for $k$ is $\rm{min}(1,-k_{-})$ from which using $x=(2k+\alpha\omega)/q$
with $\alpha=-1$, we find that $x_{\rm max}$  can be either $(2-\omega)/q$ or $-1$. 
Putting the first and second terms together we obtain,
\bearr
   F_B&=&\Theta(1-k_{+})g_B\left(\frac{2-\omega}{q}\right)+g_B\left(-\frac{\omega}{q}\right)\label{gB.eqn}\\
   &&-\Theta(1+k_{-})g_B(-1)-\Theta(-k_{-}-1)g_B\left(\frac{2-\omega}{q}\right).\nn
\eearr

Having completely calculated the inter-valley $\Delta\chi^0_{+-}$, according to 
Eq.~\eqref{twoterms.eqn} we only need to calculate $\chi^0_{u,+-}$ to complete 
the evaluation of the total inter-valley propagator $\chi^0_{+-}$ which can be done in a nice
covariant way, and is the subject of the following section.

\section{Evaluation of the undoped inter-valley polarization}
\label{interundoped.sec}
In this appendix we will use a covariant notation, so the 3-vector $k^\mu$ 
denotes $(k_0,\vk)$ where $\vk=k_x\hat x + k_y \hat y$ is the momentum in two-dimensional plane. 
So in this section $q$ is not the magnitude of $\vq$, but is rather $(q_0,\bq)$. 
As we will see shortly, in the calculation of inter-valley polarization for undoped Dirac sea,
a term will appear which is exactly the intra-valley polarization. Therefore both to adjust the notation, 
and grasp the essential mathematical steps, 
let us first reproduce the diagonal correlator $\chi^0_{u,++}(q)=-\langle G_+(k)G_+(k+q)\rangle$,
where the minus comes from the fermion loop. 

\subsection{Intra-valley term}
The Dirac Hamiltonian~\eqref{dirac.eqn} corresponds to the following choice of Dirac's gamma matrices
\be
   \gamma^0=\sigma_z,~~~\gamma^1=i\sigma_y,~~~\gamma^2=-i\tau \sigma_x.
   \label{gammas.eqn}
\ee
They satisfy the Clifford algebra $\{\gamma^\mu,\gamma^\nu\}=2\eta^{\mu\nu}$ with the convention $\eta^{\mu\nu}={\rm diag}(1,-1,-1)$
for the metric of the $2+1$-dimensional Minkowski space~\cite{SemenoffChapter}. 
Note that since we are not interested in chiral symmetry breaking, we do not augment the above $2\times 2$ 
representation to $4\times 4$ representation~\cite{Pisarski1984}. 
The crucial point is that the $\gamma$ matrices or the Hamiltonian of the {\em other valley} can be obtained
from the above representation by $\gamma^\mu\to \gamma^{\mu*}$. The identity $\gamma^{\mu*}=\gamma^2\gamma^\mu\gamma^2$~\cite{ZeeQFTBook}
connects the two representations. This essentially is the expression of the fact that the state around one valley
are obtained from the other valley by time-reversal operation~\cite{ZeeQFTBook}. 

To set the stage for the calculation of valley-off-diagonal density-density correlation function within the
covariant notation, let us first summarize the calculation of Son~\cite{Son2007}. For the intra-valley situation, 
the density-density correlation function is given by,
\bearr
   &&\chi^0_{u,++}(q)=-\int\frac{d^3k}{(2\pi)^3}\mbox{Tr}\left(\gamma^0\frac{1}{\slashed k}\gamma^0\frac{1}{\slashed k+\slashed q} \right)
   \label{looppp.eqn}
\eearr
where the covariant notation $k^\mu=(k_0,\vk)$ and $q^\mu=(q_0,\vq)$ and the Feynman slash notation $\slashed k=\gamma^\mu k_\mu$ is understood. 
Both momenta $k^\mu$ and $k^\mu+q^\mu$ belong to the same valley and hence the same representation of the Dirac matrices $\gamma^\mu$.
The Feynman propagators both belong to the same valley and hence are calculated from the same representation of the Hamiltonian as in Eq.~\eqref{dirac.eqn}
that corresponds to e.g. the choice~\eqref{gammas.eqn}.
To proceed with the evaluation of this integral, we need trace identities. 
In contrast to Ref.~\cite{Son2007} that uses the $4\times 4$ representation of Dirac $\gamma^\mu$ matrices, 
in this work we are interested in $2\times 2$ representations as in Eq.~\eqref{gammas.eqn}. 
The reason we use $2\times 2$ representation is that we are not interested in a further gamma matrix to 
anticommute with all $\gamma^\mu$'s with $\mu=0,1,2$~\cite{Pisarski1984}. 
The appropriate trace identity in this case will become~\cite{Peskin},
\be
   \mbox {tr}\left(\gamma^\mu\gamma^\rho\gamma^\nu\gamma^\sigma\right) =    
   2(\eta^{\mu\rho}\eta^{\nu\sigma}-\eta^{\mu\nu}\eta^{\rho\sigma}+\eta^{\mu\sigma}\eta^{\rho\nu}).
   \label{trace2pp.eqn}
\ee
Note that in contrast to Ref.~\cite{Son2007} the overall factor in the right hand side is $2$.
This arises from the fact that in Eq.~\eqref{gammas.eqn} we have used $2\times 2$ representation 
of the Clifford algebra~\cite{SemenoffChapter}. It is easy to understand the difference between the two 
factors: To survive the trace, we always need an even number of $\gamma$-matrices to pair up
to square to unit matrix of appropriate dimension~\cite{Arfken}. This fixes the overall factor as the
trace of the appropriate unit matrix which in the case of $2\times 2$ matrices is $2$ while for
$4\times 4$ matrices is $4$. Using the trick of Feynman parameters~\cite{Peskin},
\be
   \frac{1}{AB}=\int_0^1\frac{dx}{\left[xA+(1-x)B\right]^2},
\ee
and defining $\ell=k+q$, the integral will simplify to~\cite{Son2007}
\be
   \chi^0_{u,++}(q)=-2\int_0^1dx\int\frac{d^3k}{(2\pi)^3}
   \frac{2k_0\ell_0-k.\ell}{\left[x\ell^2+(1-x)k^2\right]^2}
\ee
Employing the standard formulas for dimensional regularization~\cite{Son2007}~\footnote{See page 251 of Peskin's book~\cite{Peskin}.} one obtains
\be
   \chi^0_{u,++}=-\frac{1}{16}\frac{\bq^2}{\sqrt{v_F^2{\bq}^2-\omega^2}}
   \label{chi0BaskaranJafari.eqn}
\ee
Inserting it in Eq.~\eqref{spinRPA.eqn} gives the spin-1 modes of undoped Dirac cone near the 
$\Gamma$-point~\cite{Jafari2002}. This verifies that the covariant evaluation and a more messy
way of doing Matsubara summation first and then the resulting $\bs k$-integration~\cite{Jafari2002,Jafari2005,Wunsch,Sarma2007}
give the same result~\footnote{Note that in the graphene literature a degeneracy factor of $2\cdot 2=4$ arising from two valleys
and two spin direction is also included. Also note that in contrast to Ref.~\cite{Son2007}, we have used 
a Minkowski signature for the norm of three-vectors.}. 

\subsection{Inter-valley term}

Now we are ready to work out the inter-valley polarization of the undoped 2+1 dimensional
Dirac fermions given by $\chi^0_{u,+-}=-\langle G_+(k)G_-(k+q)\rangle $ as the 
convolution of two Green's function from two different valleys $\tau=\pm$. This justifies the use of
two colors to draw the propagators in Fig.~\ref{diagram.fig}. 
Again the minus sign comes from the fermion loop forming the bubble. 
As pointed out in discussion of Eq.~\eqref{gammas.eqn}, the Hamiltonian of the $\tau=-1$ valley 
can be obtained from the $\tau=+1$ valley by changing Dirac matrices as $\gamma^\mu\to\gamma^{\mu*}=\gamma^2\gamma^\mu\gamma^2$. 
The Eq.~\eqref{looppp.eqn} will be replaced by
\bearr
   &&\chi^0_{u,+-}(q)=-\int\frac{d^3k}{(2\pi)^3}\mbox{Tr}\left(\gamma^0\frac{1}{\slashed k}\gamma^0\frac{1}{\bar{\slashed k}+\bar{\slashed q}} \right)
   \label{looppm.eqn}
\eearr
where the notation $\bar{\slashed k}=\gamma^{\mu *}k_\mu=\gamma^2\gamma^\mu\gamma^2 k_\mu$ emphasizes that
{\em the second propagator belongs to the other valley where the representation $\gamma^{\mu *}$ of the Clifford algebra is used.} 
In this case we will need the following trace formula
\bearr
   &&\mbox{tr}\left( \gamma^\mu\gamma^\rho\gamma^\nu\gamma^2\gamma^\sigma\gamma^2\right)=
   2(\eta^{\mu\rho}\eta^{\nu\sigma}-\eta^{\mu\nu}\eta^{\rho\sigma}+\eta^{\mu\sigma}\eta^{\rho\nu})\nn\\
   &&+4\left(\eta^{2\sigma}\eta^{\mu\rho}\eta^{\nu 2}-\eta^{2\sigma}\eta^{\mu\nu}\eta^{\rho 2}+\eta^{2\sigma}\eta^{\mu 2}\eta^{\rho \nu}\right)
   \label{trace2pm.eqn}
\eearr
This can be obtained by writing $\gamma^2\gamma^\sigma=\{\gamma^2,\gamma^\sigma\}-\gamma^\sigma\gamma^2=2\eta^{2\sigma}-\gamma^\sigma\gamma^2$.
Then using the fact that in our representation $\eta^{\alpha\beta}={\rm diag}(1,-1,-1)$ the $\gamma^2$ matrix squares to $-{\bs 1}_{2\times 2}$ matrix.  
The first line of the new trace formula~\eqref{trace2pm.eqn} is identical to Eq.~\eqref{trace2pp.eqn}. 
The second line plays the role of inter-valley overlap factors Eq.~\eqref{formfactor.eqn} for $\lambda=-\lambda'$ 
situation where $\tau\tau'=-1$ (inter-valley) form factor becomes non-zero and gives a non-zero phase space in the 
$|\bq|\to 0$ limit. 

For the density-density correlator we will need the $\mu\nu=00$ component of the above trace formula. Since the contribution from
the first term is identical to $\chi^0_{u,++}$, we only need the second term. Let us call it $\delta\chi^0_{u}$ which 
will simplify to
\be
   \delta\chi^0_{u}(q)=-4\int_0^1dx\int\frac{d^3k}{(2\pi)^3}
   \frac{k_y(k_y+q_y)}{\left[x(k+q)^2+(1-x)k^2\right]^2}.
\ee
The reason for picking the $y$-components is that the two valleys in Eq.~\eqref{dirac.eqn} are connected by a vector along the $k_y$ axis. 
The inter-valley nature of the process, inevitably introduces a preferred direction. In real graphene there are three of such directions, and
appropriate projection to $C_3$ representation has to be done at the end. 

Expanding the expression in the square brackets in the denominator gives $[k^2+2xk.q+xq^2]^2$.
Completing the square and defining $\tilde x=1-x$, the denominator of the above integral 
becomes $\ell^2+x\tilde xq^2$ which defines the new integration variable $\ell=k+xq$. Substituting 
$k=\ell-xq$ and performing a Wick rotation we end up with an Euclidean integration $d^3\ell$ 
will be spherically symmetric in 3-dimensions~\cite{Peskin}. Therefore terms that are odd in $\ell_y$ will not survive the integration, and
furthermore under the integration it is legitimate to replace $\ell_y^2\to \ell^2/3$ (due to isotropy of the loop integration), which eventually
yields
\be
   \delta\chi^0_{u}(q)=-4\int_0^1dx\int\frac{d^3\ell}{(2\pi)^3}
   \frac{\ell^2/3-x\tilde xq_y^2}{\left[\ell^2+x\tilde x q^2\right]^2}.
\ee
Again with the standard dimensional regularization formulas, and noting that
$\Gamma(1/2)=\sqrt{\pi}$, $\Gamma(-1/2)=-2\sqrt{\pi}$ and that
$$
   \int_0^1 \sqrt{x(1-x)}dx=\frac{\pi}{8},
$$
upon rotating back from Euclidean to Minkowski space (thereby $q^2\to q_0^2-\bq^2$),
and taking care of the factor $i$ left from Wick rotation, we obtain
\be
   \delta\chi^0_u(\omega,\bq)=\frac{1}{16}\frac{\omega^2-q_x^2}{\sqrt{v_F^2\bq^2-\omega^2}}
   \label{delchi0.eqn}
\ee
The above expression as expected is not rotationally symmetric in $\vq$, as the presence 
of a direction connecting the two Dirac valleys breaks this symmetry. However, in 
realistic graphene, there are three such directions connected to each other by 
three-fold rotations, $C_3$. Using the symmetric irreducible representation of this group 
to project the right-hand-side of the above equation~\cite{DresselhauseGroupTheory} amounts to
replacing
\be
   q_x^2\to \frac{q_x^2+\left(cq_x+sq_y\right)^2+\left(cq_x-sq_y\right)^2}{3}=\frac{q_x^2+q_y^2}{2}
\ee
where $c=\cos(2\pi/3)$ and $s=\sin(2\pi/3)$. 
Adding the two contributions in Eq.~\eqref{delchi0.eqn} and~\eqref{chi0BaskaranJafari.eqn} gives
the result presented in Eq.~\eqref{chi0JafariKoenig.eqn}. 

\bibliographystyle{apsrev4-1}
\bibliography{Refs}

\end{document}